\documentclass[%
 aip,
 amsmath,amssymb,
reprint, 
]{revtex4-1}

\usepackage{graphicx}
\usepackage{dcolumn}
\usepackage{bm}

\usepackage[utf8]{inputenc}
\usepackage[T1]{fontenc}
\usepackage{mathptmx}
\usepackage{etoolbox}

\makeatletter
\def\@email#1#2{%
 \endgroup
 \patchcmd{\titleblock@produce}
  {\frontmatter@RRAPformat}
  {\frontmatter@RRAPformat{\produce@RRAP{*#1\href{mailto:#2}{#2}}}\frontmatter@RRAPformat}
  {}{}
}%
\makeatother
\begin{document}

\preprint{AIP/123-QED}

\title[A vacuum-ultraviolet spectropolarimeter for an electron beam ion trap]{A vacuum-ultraviolet spectropolarimeter for an electron beam ion trap}

\author{Nobuyuki Nakamura}
 \email{n\_nakamu@ils.uec.ac.jp.}
\affiliation{ 
Institute for Laser Science, The University of Electro-Communications, Chofu, Tokyo 182-8585, Japan
}%

\author{Ryohko Ishikawa}%
\affiliation{ 
National Astronomical Observatory of Japan, Mitaka, Tokyo 181-8588, Japan
}%

\author{Motoshi Goto}
\affiliation{%
National Institute for Fusion Science, Toki, Gifu 509-5202, Japan
}%

\date{\today}

\begin{abstract}
We have developed a vacuum-ultraviolet spectropolarimeter for an electron beam ion trap (EBIT) to measure the linear polarization of emission lines from multiply charged ions around the Lyman-$\alpha$ wavelength.
The main components for polarimetry are a rotatable MgF$_2$ waveplate and a SiO$_2$/MgF$_2$ multilayer-coated fused silica plate that functions as a reflective polarizer.
A grazing-incidence grating is mounted between them to provide wavelength dispersion.
The polarization is determined from the intensity modulation of the spectral line as the waveplate is rotated.
The performance of the spectropolarimeter was demonstrated by measuring the polarization of the $2s$--$2p_{3/2}$ transition in Li-like N$^{4+}$ (124~nm) excited by a 1000~eV electron beam in an EBIT.
Clear modulation of the line intensity was observed as a function of the waveplate rotation angle.
From the measured modulation amplitude, the degree of linear polarization was determined to be $P=-(0.178^{+0.014}_{-0.005})$, with the negative sign indicating that the emission is polarized predominantly perpendicular to the electron beam.
This result demonstrates the capability of the present spectropolarimeter to determine polarizations with an absolute uncertainty $\Delta P$ on the order of $0.01$.
This instrument provides a useful tool for benchmarking magnetic-sublevel-resolved collision theories through polarization measurements of multiply charged ions excited by a unidirectional electron beam.

\end{abstract}

\maketitle

\section{\label{sec:intro}Introduction}

Polarization is one of the fundamental properties of radiation and provides important information on the physical processes that generate the radiation \cite{Fano3,Fujimoto2}.
The polarization of emitted radiation generally originates from anisotropy in the radiation source or its environment.
Therefore, measurements of polarization provide a powerful diagnostic of such anisotropic conditions.
For example, polarization measurements have been used to diagnose anisotropic electron velocity distributions in plasmas \cite{Fujimoto1,Henoux1,FUjimoto3,Goto2,Goto3,Ramaiya1} and to infer magnetic field structures in astrophysical and laboratory environments \cite{Deglinnocenti1,Trujillobueno1}.
In atomic radiation, the polarization is determined by the population distribution among magnetic sublevels of the excited state.
Thus, measurements of polarization allow investigations of excitation processes at the level of magnetic sublevels, providing more detailed information than intensity measurements alone \cite{Balashov1}.

For the diagnostics described above, the mechanisms responsible for generating the polarization must be well understood.
For example, interpretation of polarization measurements in terms of anisotropic electron velocity distributions requires knowledge of the magnetic-sublevel population distribution produced by excitation by a directional electron beam.
In most applications, these population distributions are obtained from theoretical calculations.
Therefore, experimental validation of the underlying theory is essential for establishing the reliability of polarization-based diagnostics.

An electron beam ion trap (EBIT) \cite{Marrs1,yebisu_nakamura} provides a well-controlled environment in which multiply charged ions are excited by a nearly monoenergetic and unidirectional electron beam.
Such conditions make EBITs particularly suitable for benchmarking theoretical predictions of magnetic-sublevel populations and the resulting polarization.
However, experimental polarization data for highly charged ions remain scarce, especially in the vacuum ultraviolet (VUV) and extreme ultraviolet (EUV) regions.
The present work aims to provide benchmark data for detailed collision theories that resolve the populations of individual magnetic sublevels and predict the resulting polarization, thereby supporting the development of reliable polarization diagnostics for laboratory and astrophysical plasmas.

In the visible wavelength region, well-developed polarimetric techniques exist because high-quality polarizers are readily available and inexpensive.
In the X-ray region, several established techniques are used depending on the photon energy, such as Bragg reflection polarimetry\cite{Henderson1,Beiersdorfer6}, photoelectron angular distribution analysis \cite{Costa1,Iwata1}, and Compton polarimetry \cite{Shah1,Nakamura34}.
In contrast, in the VUV and EUV regions, reliable and widely established techniques for polarization measurements are limited.
This is mainly because conventional transmissive polarizers are not available in these wavelength ranges and most materials exhibit strong absorption, which restricts the choice of optical elements for polarization analysis.
Furthermore, polarization-sensitive interactions commonly exploited at higher photon energies are generally less practical in the VUV/EUV region, making accurate polarization measurements technically challenging.

In this work, we have developed a spectropolarimeter capable of measuring the polarization of radiation in the VUV region.
The design of the instrument incorporates key technologies developed for the sounding rocket experiment Chromospheric Lyman-Alpha Spectro-Polarimeter (CLASP) \cite{Kano1,Watanabe19}, which aimed at measuring the polarization of the solar Lyman-$\alpha$ radiation originating from the solar chromosphere and the transition region.
The spectropolarimeter was installed on a compact EBIT \cite{cobit} to test its performance.
As a demonstration experiment, we measured the polarization of the $2s$--$2p_{3/2}$ transition in Li-like N$^{4+}$ excited by an electron beam.

\section{\label{sec:design}Spectropolarimeter design}
\begin{figure}
\includegraphics[width=0.48\textwidth]{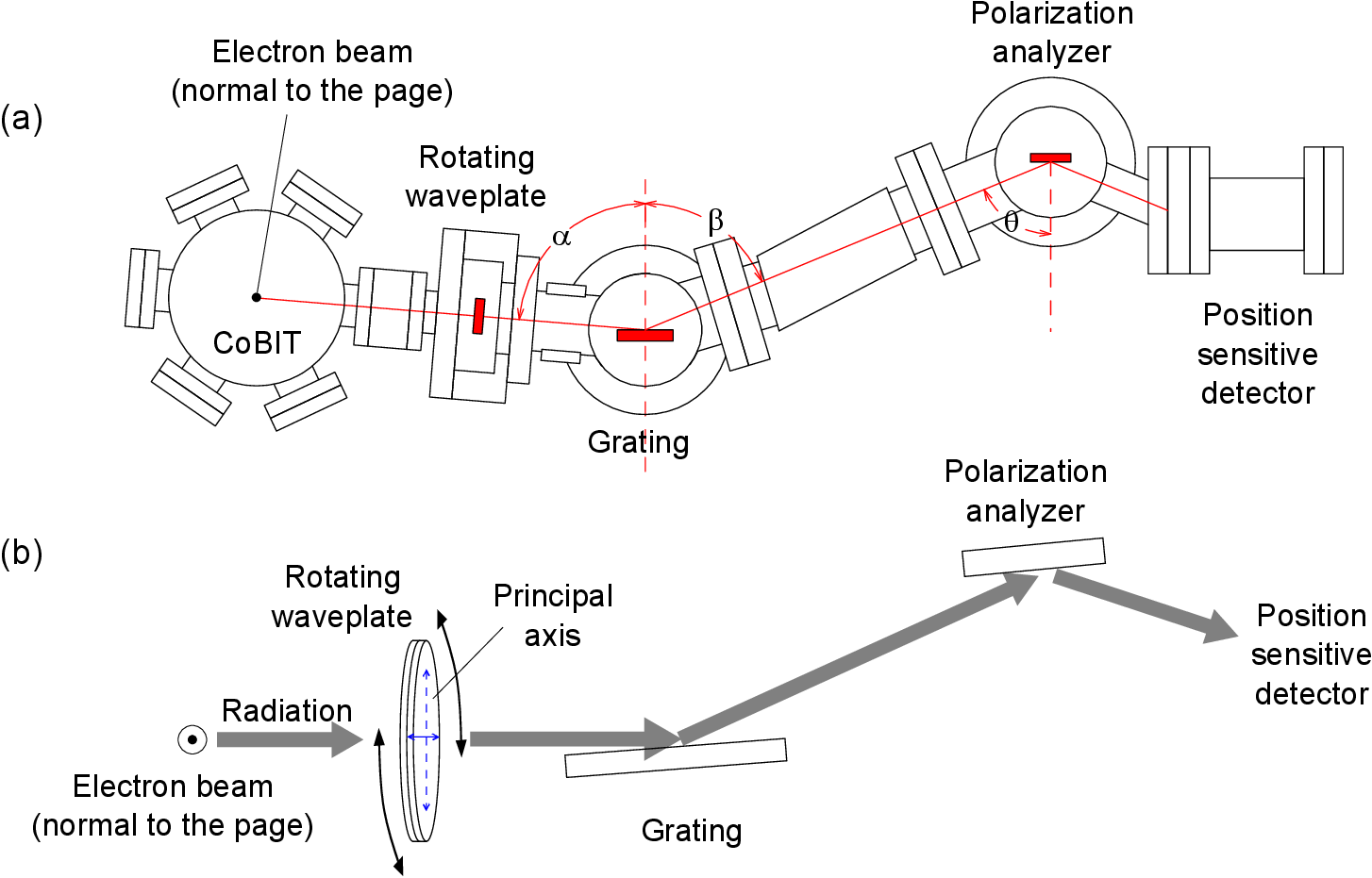}
\caption{\label{fig:setup} (a) Schematic diagram of the present spectropolarimeter coupled to the compact electron beam ion trap (CoBIT).
The optical components consist of a rotating waveplate, a grazing-incidence grating, a polarization analyzer, and a position-sensitive detector.
(b) Measurement geometry.
The electron beam propagates normal to the plane of the page, defining the quantization axis. The radiation is observed at 90° with respect to the beam axis.
The polarization analyzer preferentially reflects radiation polarized parallel to the electron beam axis in this geometry.}
\end{figure}

The present spectropolarimeter was developed by modifying a grazing-incidence flat-field spectrometer reported previously \cite{Nakamura31}.
Figure~\ref{fig:setup} shows a schematic drawing of the instrument.
For the performance test described in Sec.~\ref{sec:performance}, the spectropolarimeter was coupled to a compact EBIT, CoBIT \cite{cobit}.
In the present study, we focus on linear polarization measurements of radiation observed at 90$^\circ$ relative to a well-defined quantization axis, where the linear polarization $P$ is defined as:
\begin{equation}
P =
\frac{I_\parallel - I_\perp}
     {I_\parallel + I_\perp},
\end{equation}
where $I_\parallel$ and $I_\perp$ represent the intensities of radiation with polarization vectors parallel and perpendicular to the quantization axis, respectively.

\subsection{\label{sec:principle}Principle}

The present spectropolarimeter is based on the technique used in the rocket experiment CLASP \cite{Kano1}.
The instrument consists of a waveplate, a diffraction grating, a polarization analyzer, and a detector.
The waveplate changes the polarization state of the incident radiation, the diffraction grating provides wavelength dispersion, and the polarization analyzer preferentially reflects one polarization component.
As a result, the photon counts detected by the detector exhibit a periodic modulation as a function of the waveplate rotation angle.
The amplitude of this modulation is proportional to the degree of linear polarization of the radiation.
From this modulation, the polarization properties of the emitted radiation can be determined.

The observed photon intensity $I^\mathrm{obs}(\phi)$ at the detector can be expressed as a function of the waveplate rotation angle $\phi$ as \cite{Ishikawa1}:
\begin{widetext}
\begin{equation}
I^\mathrm{obs}(\phi) =
\frac{I}{2}
\left[
\left( R_s+ R_p \right)
+
P \left( R_s - R_p \right)
\left\{
\frac{1+\cos \delta}{2}
+
\frac{1-\cos \delta}{2}
\cos \left[ 4 \left( \phi - \phi_0 \right) \right]
\right\}
\right],
\label{eq:intensity}
\end{equation}
\end{widetext}
where $I$ is the radiation intensity, $\delta$ is the retardation of the waveplate, and $\phi_0$ is the angle between the principal axis of the waveplate and the quantization axis.
$R_s$ and $R_p$ represent the efficiencies for the $\sigma$- and $\pi$-polarized radiation, respectively.
They are given by the products of the diffraction efficiencies of the grating ($R^\mathrm{g}_s$ and $R^\mathrm{g}_p$) and the reflectivities of the polarization analyzer ($R^\mathrm{p}_s$ and $R^\mathrm{p}_p$), i.e., $R_s = R^\mathrm{g}_s R^\mathrm{p}_s$ and $R_p = R^\mathrm{g}_p R^\mathrm{p}_p$.

From Eq.~(\ref{eq:intensity}), the observed intensity exhibits a modulation of the form
\begin{equation}
I^\mathrm{obs}(\phi) =
A
\left(
1 + B \cos \left[ 4 \left( \phi - \phi_0 \right) \right]
\right),
\label{eq:mod1}
\end{equation}
where
\begin{eqnarray}
A &=&
\frac{I}{2}
\left( R_s+ R_p \right)
\left(
1+ P' \frac{1+\cos \delta}{2}
\right)
\\
B &=&
\frac{P'(1- \cos \delta)}
     {2+P' (1+\cos \delta)}
\label{eq:B}
\\
P'&=&
P
\frac{R_s - R_p}{R_s + R_p}.
\label{eq:P'}
\end{eqnarray}
The magnitude of the parameter $B$ represents the modulation amplitude.
The parameter $B$ becomes identical to $P'$ when the waveplate is an ideal $\lambda/2$ plate, i.e., $\delta = 180^\circ$.
Furthermore, when the polarizing power, defined by $(R_s - R_p)/(R_s + R_p)$, is unity, the parameter $B$ directly gives the polarization $P$.
Under the condition $(R_s - R_p)/(R_s + R_p) > 0$ (i.e., $R_s > R_p$), the parameter $B$ and the polarization $P$ take the same sign.
Therefore, a positive $P$ (and thus a positive $B$) yields a modulation maximum at $\phi=\phi_0$, and vice versa.

\subsection{\label{sec:component}Optical components}

The waveplate \cite{Ishikawa1} is a zero-order type constructed by stacking two MgF$_2$ plates in optical contact, with their principal axes oriented orthogonally.
These two plates differ slightly in thickness, and the retardation $\delta$ at wavelength $\lambda$ is determined by the thickness difference $\Delta d$ and the birefringence $n_e-n_o$ at $\lambda$ as
\begin{equation}\label{eq:retardation}
\delta = 2\pi \frac{(n_e-n_o)\Delta d}{\lambda}.
\end{equation}
The waveplate functions as a $\lambda/2$ plate when $\delta = 180^\circ$ (or $180^\circ \times M$ where $M=0,\pm1,\pm2,...$).
The waveplate is mounted on an ultra-high-vacuum rotation stage (SmarAct SR-5714C) to control the waveplate rotation angle.
We have several waveplates with different $\Delta d$ values that were produced during the development of CLASP.
Their retardations were calibrated at UVSOR and confirmed to satisfy $\delta = 180^\circ \times M$ within the wavelength range of 114--126~nm \cite{Ishikawa1}.
In the present study, two waveplates with $\Delta d = 8.420\pm0.050~\mu$m and $14.495\pm0.050~\mu$m were used.
For the waveplate with $\Delta d = 8.420~\mu$m, the retardation is close to $180^\circ$ at approximately 124~nm, corresponding to the wavelength of the Li-like N$^{4+}$ transition investigated in this work.

The polarization analyzer \cite{Narukage1} is a fused silica plate coated with alternating SiO$_2$ and MgF$_2$ layers and designed to function as a reflective polarizer for Lyman-$\alpha$ radiation at a Brewster angle of approximately $68^\circ$.
A polarizing power $(R_s^\mathrm{p}-R_p^\mathrm{p})/(R_s^\mathrm{p}+R_p^\mathrm{p})$ of 0.99 has been confirmed at the Lyman-$\alpha$ wavelength (121.57~nm) \cite{Narukage1}.
In the present setup, the incident angle is set to $\theta=67.5^\circ$.

The grating and the position-sensitive detector (PSD) are the same components used in the original spectrometer configuration \cite{Nakamura31}.
The grating (Hitachi High-Tech Corp., 001-0639) is an aberration-corrected variable-line-spacing type that enables flat-field focusing.
The groove density is 600~mm$^{-1}$, and the working incidence angle $\alpha$ is $85.3^\circ$.
The PSD (Quantar 3391) consists of five microchannel plates (MCPs) and a resistive anode.
The surface of the front MCP is coated with CsI to enhance the detection efficiency for photons in the VUV range.

\section{\label{sec:performance}Performance Measurements}

The performance of the spectropolarimeter was evaluated by observing the $2s$--$2p_{3/2}$ transition in Li-like N$^{4+}$ at 123.88~nm using CoBIT \cite{cobit}.
CoBIT consists of an electron gun, an ion trap, an electron collector, and a superconducting magnet.
The ion trap is composed of three successive cylindrical electrodes that provide an axial potential well for trapping ions.
Radial confinement is achieved by the space-charge potential of the electron beam, emitted from the gun and compressed by the magnetic field of the superconducting magnet, as it travels through the ion trap.
Multiply charged ions are thus produced through successive electron-impact ionization.
To produce Li-like N$^{4+}$, nitrogen gas was introduced through a variable leak valve connected to one of the side ports.
Radiation from the trapped N$^{4+}$ ions, excited by the electron beam, was observed using the spectropolarimeter positioned at a 90$^\circ$ angle relative to the beam axis (see Fig.~\ref{fig:setup}).
Due to the cylindrical symmetry of the collision system, the electron beam direction defines the quantization axis.
It should be noted that, in the present geometry, the polarization component parallel to the electron beam corresponds to the $\sigma$ polarization for the grating and the polarization analyzer.

The electron beam energy was set to 1000~eV, at which the degree of linear polarization predicted using the flexible atomic code (FAC) \cite{FAC} is -0.20, while the electron beam current was 13~mA.
To suppress the accumulation of unwanted ions, such as Ba and W evaporated from the electron-gun cathode, the trap was periodically emptied with a period of 0.5~s.


The intensity of the $2s$--$2p_{3/2}$ transition was recorded while rotating the waveplate over a range of $90^\circ$ in 12 steps (i.e., $7.5^\circ$ per step).
Since several hours of data acquisition were required to obtain sufficient statistics, the waveplate rotation angle was randomly changed every 1 min among the 12 predefined angles.
The spectra obtained at the same angle were subsequently summed and analyzed together.
This procedure minimizes the influence of long-term instrumental drifts by distributing measurements at each angle over the entire acquisition period. The total acquisition times were approximately 8 h and 12 h for the measurements with the $8.420~\mu$m and $14.495~\mu$m waveplates, respectively, corresponding to 480 and 720 one-minute acquisition frames in total (about 40 and 60 frames per angle, respectively, on average).

\section{Results}

\begin{figure}
\includegraphics[width=0.40\textwidth]{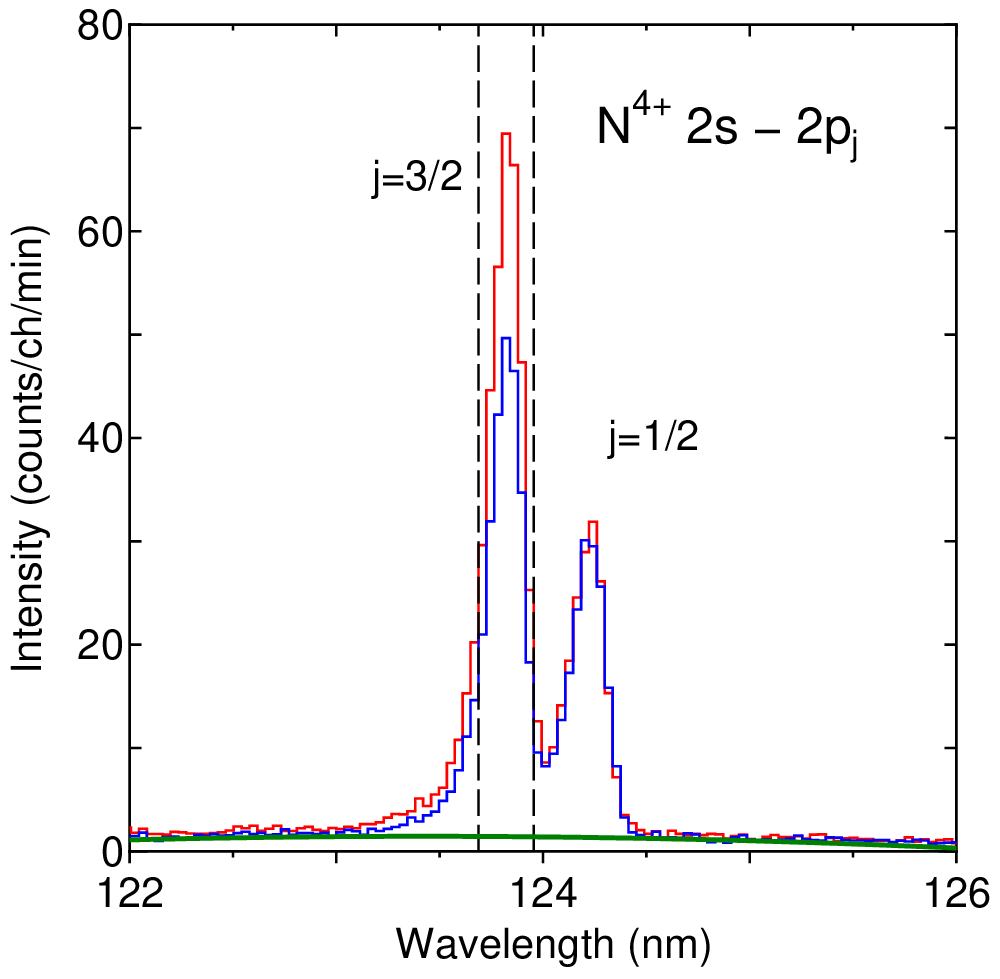}
\caption{\label{fig:spectra}
Spectra of the $2s$--$2p_j$ transitions in Li-like N$^{4+}$ observed with the present spectropolarimeter installed on a compact EBIT.
Blue and red curves represent spectra obtained at waveplate rotation angles of $\phi=7.5^\circ$ and $52.5^\circ$, respectively.
The vertical scale is normalized by the acquisition time.}
\end{figure}

Figure~\ref{fig:spectra} shows the spectra of Li-like N$^{4+}$ obtained with the present setup using the $\Delta d=8.420~\mu$m waveplate.
The peak on the shorter-wavelength side corresponds to the $2s$--$2p_{3/2}$ transition, whereas that on the longer-wavelength side corresponds to the $2s$--$2p_{1/2}$ transition.
The blue and red curves represent spectra obtained at waveplate rotation angles of $\phi=7.5^\circ$ and $52.5^\circ$, respectively.
Here, $\phi$ is the angle measured from the reference position defined by the rotation stage.
Note that at $\phi$=0, the principal axis of the waveplate has an initial offset angle ($\phi_0$ in Eq.~(\ref{eq:mod1})) with respect to the electron beam direction (i.e., the quantization axis).
As will be shown later, the offset is $\phi_0 \simeq$5$^\circ$; thus, the angles $\phi$=7.5$^\circ$ and 52.5$^\circ$ are close to $\phi_0$ and $\phi_0$+45$^\circ$, respectively, where the modulation is expected to exhibit extrema.
Since the $J=1/2$--$1/2$ transition is intrinsically isotropic and unpolarized, the intensity of the $2s$--$2p_{1/2}$ transition shows no dependence on the waveplate rotation angle, as expected.
In contrast, the $2s$--$2p_{3/2}$ transition exhibits a strong dependence, as seen in the figure.

\begin{figure}
\includegraphics[width=0.40\textwidth]{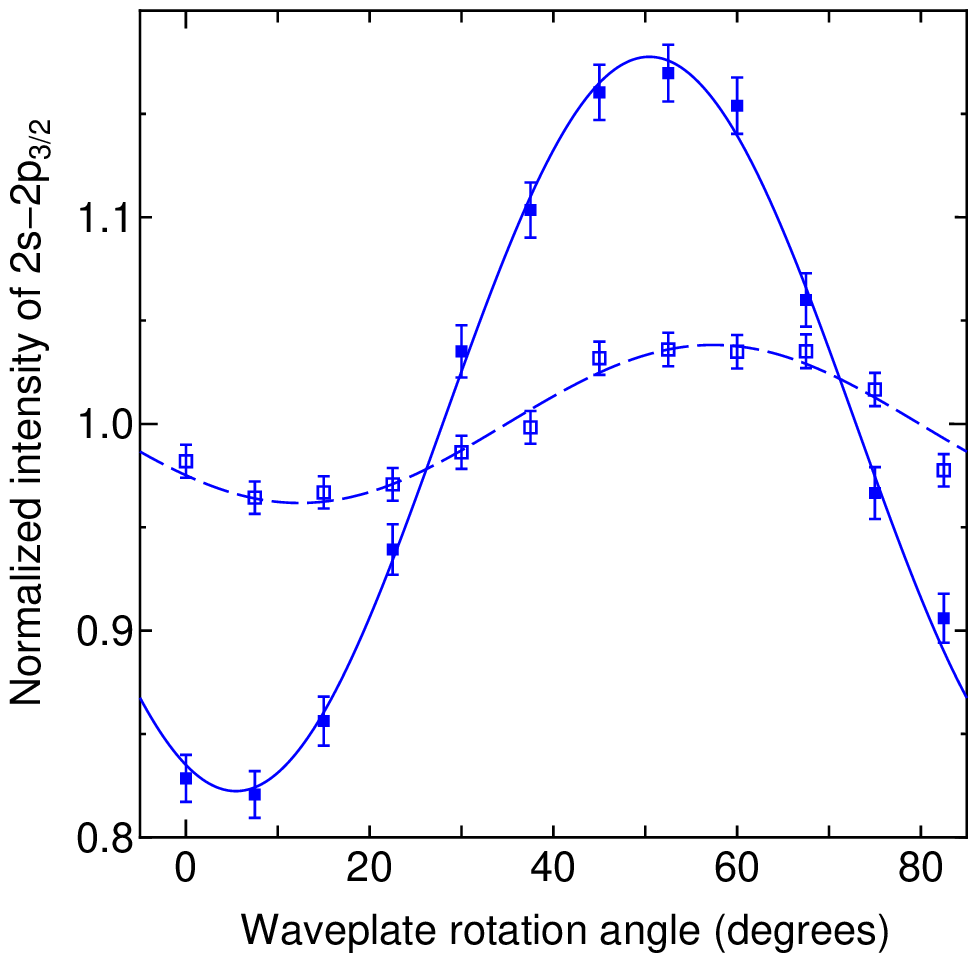}
\caption{\label{fig:modulation}
Intensity of the $2s$--$2p_{3/2}$ transition in Li-like N$^{4+}$ observed as a function of the waveplate rotation angle.
The solid and open squares represent the results obtained with the $\Delta d=8.420~\mu$m and $14.495~\mu$m waveplates, respectively.
The solid and dashed lines represent fits of Eq.~(\ref{eq:mod1}) to the respective datasets.}
\end{figure}

Figure~\ref{fig:modulation} shows the observed intensity of the $2s$--$2p_{3/2}$ transition as a function of the waveplate rotation angle.
The intensity was obtained by summing the counts over the interval bounded by the dotted lines in Fig.~\ref{fig:spectra} and subtracting the background contribution.
The background was estimated from the regions on either side of the $2s$--$2p_j$ transitions and is shown as the green line in Fig.~\ref{fig:spectra}.
The results obtained with the $8.420~\mu$m and $14.495~\mu$m waveplates are shown by the solid and open squares, respectively.
The error bars represent statistical uncertainties.
The solid and dashed lines represent fits of Eq.~(\ref{eq:mod1}) to the experimental data, where the statistical uncertainties were used as weights.
The vertical axis is normalized by the value of $A$ obtained from the fitting for each dataset.
As already explained, due to the initial offset of the waveplate's principal axis at $\phi$=0, the modulation exhibits an extremum at $\phi \neq 0$.
In addition, the orientation of the principal axis differed between the two waveplates, resulting in extrema at different $\phi$ values for the two datasets.

The parameter $B$ obtained from the fitting is $-0.178\pm0.005$ and $-0.038\pm0.003$ for the $\Delta d=8.420~\mu$m and $14.495~\mu$m data, respectively.
The waveplate was installed to roughly align its principal axis with the electron beam axis at $\phi \simeq 0^\circ$.
Consequently, the result shown in Fig.~\ref{fig:modulation} indicates that $\phi_0 \simeq 5^\circ$ rather than $50^\circ$, which corresponds to the modulation exhibiting a minimum at $\phi=\phi_0$, i.e., when the principal axis of the waveplate coincides with the electron beam direction (the quantization axis).
The parameter $B$ is therefore negative, which corresponds to a negative $P$ value.

\section{Discussion}

To deduce the polarization $P$ from the fitted $B$ parameter through Eq.~(\ref{eq:B}) and (\ref{eq:P'}), the polarizing power $(R_s-R_p)/(R_s+R_p)$ and the retardation $\delta$ at the target wavelength of 123.88~nm must be known.
In this section, we discuss the values of $(R_s-R_p)/(R_s+R_p)$ and $\delta$, and evaluate the uncertainty with which the polarization can be determined using the present spectropolarimeter.

\subsection{Porlarizing power}

\begin{figure}
\includegraphics[width=0.45\textwidth]{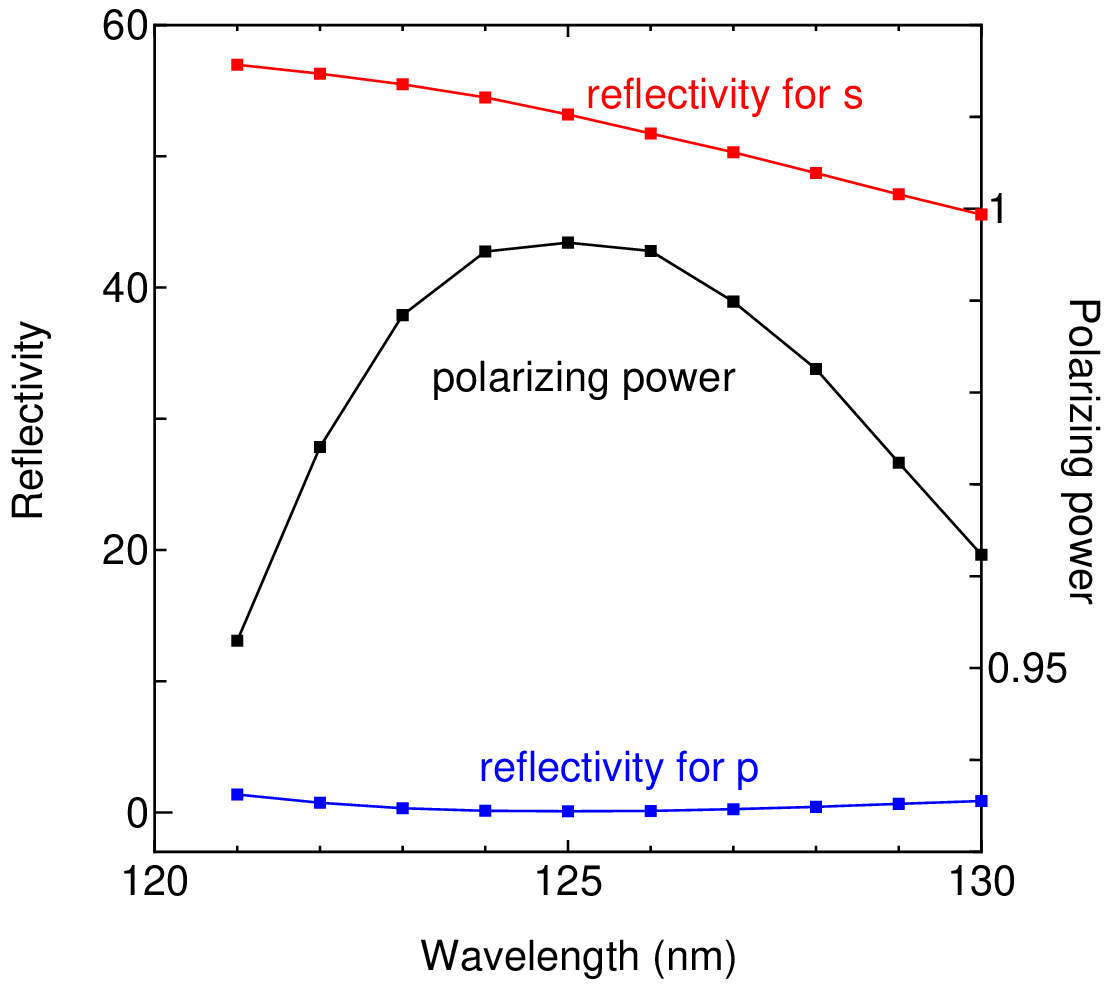}
\caption{\label{fig:reflectivity}
Reflectivities $R_s^\mathrm{p}$ and $R_p^\mathrm{p}$ of the polarization analyzer for the $\sigma$- (red) and $\pi$-polarized radiation (blue), respectively (left axis).
The polarizing power $(R_s^\mathrm{p}-R_p^\mathrm{p})/(R_s^\mathrm{p}+R_p^\mathrm{p})$ is shown in black (right axis).
}
\end{figure}

The reflectivities $R^\mathrm{p}_s$ and $R^\mathrm{p}_p$, and the polarizing power $(R^\mathrm{p}_s-R^\mathrm{p}_p)/(R^\mathrm{p}_s+R^\mathrm{p}_p)$ of a SiO$_2$/MgF$_2$ multilayer-coated fused silica plate fabricated under the same conditions and processes as those used for the polarization analyzer in the present instrument were measured by Narukage \textit{et al.} \cite{Narukage1} and Goto \textit{et al.} \cite{Goto1}.
Figure~\ref{fig:reflectivity} shows the wavelength dependence of the reflectivities and the polarizing power at an incident angle of $68^\circ$ measured by Goto \textit{et al.}
As shown in the figure, the polarizing power reaches approximately 0.99 at 123.88~nm.

The incident angle in the present instrument is determined by the mechanical accuracy of the mounting and is estimated to be $67.5\pm2.0^\circ$.
Although the incident-angle dependence of the polarizing power has not been measured at 123.88~nm, measurements at the Ly-$\alpha$ wavelength reported by Narukage \textit{et al.} \cite{Narukage1} suggest that the polarizing power may decrease by approximately 0.01--0.02 for an incident-angle variation of $\pm2.0^\circ$.
Thus, the polarizing power of the polarization analyzer, $(R^\mathrm{p}_s-R^\mathrm{p}_p)/(R^\mathrm{p}_s+R^\mathrm{p}_p)$, is estimated to be in the range 0.97--0.99.

The polarization dependence of the diffraction efficiency of the grating must also be considered when evaluating the total polarizing power of the instrument,
\begin{equation}
\frac{R_s-R_p}{R_s+R_p}
=
\frac{R^\mathrm{g}_s R^\mathrm{p}_s - R^\mathrm{g}_p R^\mathrm{p}_p}
     {R^\mathrm{g}_s R^\mathrm{p}_s + R^\mathrm{g}_p R^\mathrm{p}_p}.
\end{equation}
Although $R^\mathrm{g}_s$ and $R^\mathrm{g}_p$ are not known for the grating used in the present instrument, it is generally known that $R^\mathrm{g}_s$ is larger than $R^\mathrm{g}_p$ for grazing-incidence gratings (see, for example, Ref.~\onlinecite{Arakawa1}).
Therefore, the total polarizing power of the instrument should not be smaller than that of the polarization analyzer alone.
Thus, $(R_s-R_p)/(R_s+R_p)$ is expected to be larger than 0.97.

\subsection{Retardation}

\begin{figure}
\includegraphics[width=0.40\textwidth]{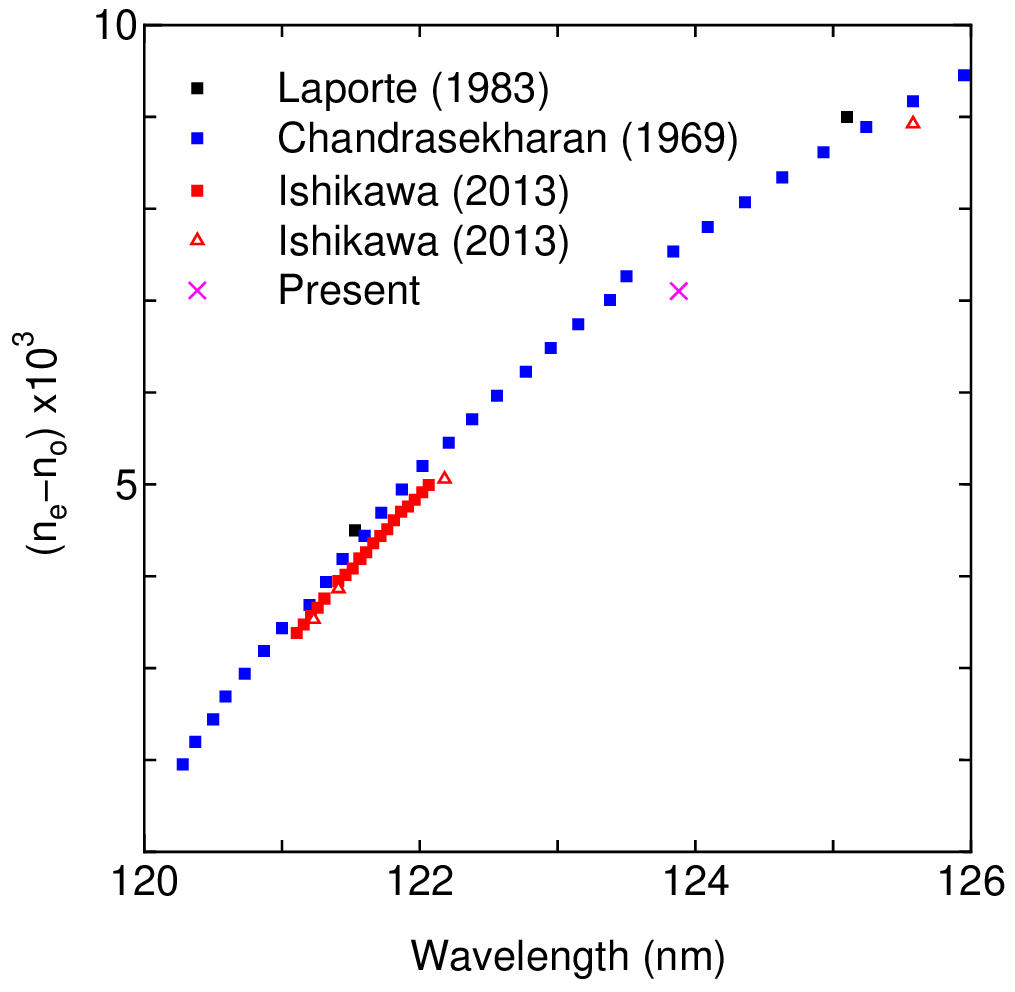}
\caption{\label{fig:birefringence}
Birefringence $(n_e-n_o)$ of MgF$_2$ reported by Laporte \cite{Laporte1}, Chandrasekharan and Damany \cite{Chandrasekharan1}, and Ishikawa \cite{Ishikawa1}.
The red squares and triangles represent the values taken from Fig.~4 and Fig.~2 of Ref.~\onlinecite{Ishikawa1}, respectively.
The value estimated from the present measurements using two waveplates with different $\Delta d$ values is shown as a magenta cross.}
\end{figure}

The retardation $\delta$ of the waveplate is determined from the birefringence $(n_e-n_o)$ and the thickness difference $\Delta d$ as shown in Eq.~(\ref{eq:retardation}).
Figure~\ref{fig:birefringence} shows the wavelength dependence of the birefringence reported in the literature.

The blue squares represent the data reported by Chandrasekharan and Damany \cite{Chandrasekharan1}.
According to their results, the birefringence at 123.88~nm is approximately $7.58\times10^{-3}$.
Using this value, the retardation $\delta$ is calculated to be $320^\circ$ and $184^\circ$ for the $\Delta d=14.495~\mu$m and $8.420~\mu$m waveplates, respectively.

The birefringence values obtained by Laporte \textit{et al.} \cite{Laporte1} are also plotted as black squares in Fig.~\ref{fig:birefringence}.
Although only two data points are available in this wavelength region, the Laporte values are slightly larger than those of Chandrasekharan and Damany by about 2\%.

On the other hand, Ishikawa \textit{et al.} \cite{Ishikawa1} accurately determined the birefringence around the Lyman-$\alpha$ wavelength.
The red squares represent their data taken from Fig.~4 of Ref.~\onlinecite{Ishikawa1}.
In addition, the wavelength at which the retardation becomes $180^\circ$ was measured for several waveplates with different $\Delta d$ values.
From these measurements (Fig.~2 of Ref.~\onlinecite{Ishikawa1}), the birefringence at several wavelengths can be deduced.
Those values are also plotted in Fig.~\ref{fig:birefringence} as red triangles.
As shown in the figure, the birefringence values derived from Ref.~\onlinecite{Ishikawa1} are generally smaller than those reported by Chandrasekharan and Damany \cite{Chandrasekharan1}, with a difference of about 3\%.

The birefringence at 123.88~nm can also be estimated from the present measurements by assuming that the polarization $P$, and therefore $P'$, was the same for the measurements using the two waveplates because the experimental conditions were essentially the same.
Experimentally, $B=-0.038$ for $\Delta d=14.495~\mu$m and $B=-0.178$ for $\Delta d=8.420~\mu$m were obtained.
Using Eqs.~(\ref{eq:B}) and (\ref{eq:retardation}), the birefringence $(n_e-n_o)$ is derived to be $7.1\times10^{-3}$ if both measurements correspond to radiation with the same $P$ (or $P'$).
This value is plotted in Fig.~\ref{fig:birefringence} as the magenta cross.
It is about 6\% smaller than the value reported by Chandrasekharan and Damany \cite{Chandrasekharan1}.

Assuming that the birefringence lies in the range from $-6\%$ to $+2\%$ relative to the value reported by Chandrasekharan and Damany, i.e., $(7.13$--$7.73)\times10^{-3}$, and considering that $\Delta d$ has an uncertainty of $0.05~\mu$m, the retardation $\delta$ of the $\Delta d=8.420~\mu$m waveplate is estimated to be in the range $173^\circ$--$190^\circ$.
If the retardation is $190^\circ$, corresponding to a $10^\circ$ deviation from the ideal $\lambda/2$ condition, $B$ becomes smaller than $P$ (or $P'$) by about 0.002.

\subsection{Contribution from the unpolarized $j=1/2$ peak}

As shown in Fig.~\ref{fig:spectra}, the $j=3/2$ peak of interest is not completely resolved from the neighboring $j=1/2$ peak.
Since the $j=1/2$ peak is intrinsically unpolarized, contamination of the integration region by the $j=1/2$ peak would reduce the measured degree of polarization of the $j=3/2$ transition.
To estimate the sensitivity of the result to contamination from the neighboring $j=1/2$ peak, the analysis was repeated using an integration region shifted by one channel (approximately 0.04 nm) farther away from the $j=1/2$ peak.
The resulting fitted parameter was $B=-0.180\pm0.005$, differing by about 0.002 from the value obtained using the original integration region.
Although this difference is smaller than the statistical fitting uncertainty, we conservatively regard it as a systematic uncertainty arising from contamination by the unpolarized $j=1/2$ peak.

\subsection{Final Result and Uncertainty Analysis}

Assuming $\delta = 180^\circ$ and $(R_s-R_p)/(R_s+R_p)=1$ for the $\Delta d = 8.420~\mu$m waveplate, the $B$ value directly gives the polarization; thus $P=-0.178$ with a statistical uncertainty of $\pm0.005$.
Considering that the minimum value of $(R_s-R_p)/(R_s+R_p)$ is 0.97, the magnitude of the polarization may exceed that of the measured $B$ value by at most 0.005.
The uncertainty in the retardation may also reduce the magnitude of the $B$ value by at most 0.002.
In addition, contamination from the neighboring $j=1/2$ peak may reduce the magnitude of the measured $B$ value by 0.002.
Thus, if these systematic uncertainties are added linearly, the final result is obtained as
\[
P=-(0.178^{+0.014}_{-0.005})
\]
for the $2s$--$2p_{3/2}$ transition in Li-like N$^{4+}$ excited by a 1000~eV electron beam.

The polarization obtained in the present study is slightly smaller in magnitude than the prediction of the flexible atomic code (FAC), which is -0.20.
This difference may be attributed to processes that are not included in the calculation.
In the calculation, only the excitation processes of Li-like N$^{4+}$ were considered, and contributions from recombination and charge-exchange processes involving He-like N$^{5+}$ were not included.
Although recombination of 1000~eV electrons by He-like N$^{5+}$ is expected to have a negligible contribution, electron capture by He-like N$^{5+}$ from residual or introduced N$_2$ gas followed by cascading may contribute to the population of the $2p_{3/2}$ level of Li-like N$^{4+}$.
A rough estimate can be made by assuming a charge-exchange cross section of $\sim10^{-15}$~cm$^2$ and a gas pressure of $\sim10^{-6}$~Pa.
Under these assumptions, the charge-exchange rate is approximately two orders of magnitude lower than the electron-impact excitation rate.
Therefore, the contribution of charge exchange to the population of the $2p_{3/2}$ level is expected to be small.
Nevertheless, since such processes would tend to reduce the magnitude of the observed polarization, their contribution may need to be considered in future studies.
It should be noted that the EBIT plasma is optically thin. Using a typical ion density of $10^{8}$--$10^{9}$ cm$^{-3}$, a plasma size of $\sim0.1$ cm, and a resonant absorption cross section of $10^{-13}$--$10^{-14}$ cm$^{2}$, the optical depth is estimated to be $\sim 10^{-6}$--$10^{-4}$.
Therefore, scattering and radiative-transfer effects, which can contribute to polarization in optically thick astrophysical plasmas, are expected to be negligible under the present experimental conditions.

\section{Summary and outlook}

We have developed a spectropolarimeter for measuring the linear polarization of vacuum-ultraviolet emission lines around the Lyman-$\alpha$ wavelength and demonstrated its performance using radiation from Li-like N$^{4+}$ ions produced in an electron beam ion trap.
The polarization of the $2s$--$2p_{3/2}$ transition excited by 1000~eV electrons was successfully determined from the intensity modulation of the spectral line induced by rotation of a MgF$_2$ waveplate.
The measured polarization was $P=-(0.178^{+0.014}_{-0.005})$, which is somewhat smaller in magnitude than the theoretical prediction obtained with the flexible atomic code.
The present study demonstrates that the developed spectropolarimeter can determine the degree of linear polarization with an absolute uncertainty on the order of 0.01.
This capability makes the instrument suitable for polarization diagnostics of vacuum-ultraviolet emission lines from laboratory plasmas, where polarization signals are often modest but contain important information on excitation processes and anisotropy of the emitting ions.

Because the thickness difference $\Delta d$ can be controlled with high precision during fabrication, waveplates optimized for other wavelengths can be designed using the available birefringence data of MgF$_2$. Therefore, the present spectropolarimeter can be applied over the wavelength range for which reliable birefringence data \cite{Ishikawa1,Laporte1,Chandrasekharan1} are available, approximately 115--135~nm. This wavelength range includes numerous VUV transitions of highly charged ions that can be produced in EBITs. Future measurements of these transitions will provide benchmark data for detailed collision theories that resolve magnetic-sublevel populations and predict the resulting polarization.

\begin{acknowledgments}
This work was supported by JSPS KAKENHI Grant Numbers JP24H00200 and the UEC-NAOJ matching fund project.
\end{acknowledgments}

\section*{AUTHOR DECLARATIONS}
\subsection*{Conflict of Interest}
The authors have no conflicts to disclose.

\section*{Data Availability Statement}
The data that support the findings of this study are available from the corresponding author upon reasonable request.


%

\end{document}